# Malus' Law for a Real Polarizer

**Ioan Damian**
University "Politehnica" Timisoara, Romania
ijdamian@yahoo.com

## 1. Introducere

In a natural light beam the electric field vector (or light vector) oscillates with the same amplitude in all planes which contain the direction of propagation. When the natural light passes through optical devices, the oscillations of the light vector in some directions can be attenuated and so we obtain the partial polarized light, like in fig. 1(in which the direction of propagation of the light beam is perpendicular to the figure plane). The degree of polarization of the light is defined as:

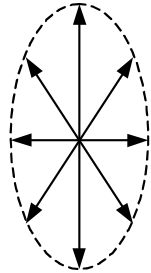
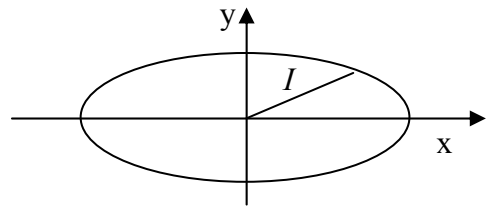

Fig. 1  Fig. 2

$$P = \frac{I_1 - I_2}{I_1 + I_2} \qquad (1)$$

where $I_1$ is the intensity corresponding to the oscillation, which has maximum amplitude, and $I_2$ is the intensity corresponding to minimum amplitude oscillation. The magnitude $P$ characterizes also the polarizing capacity of the device.

The devices called generally *polarizers* should have to produce total polarized light, i.e. the light in which vibration exists in a single plane which contains the direction of propagation. But a real polarizer, especially those that are inexpensive, furnishes rather partial polarized light. For this reason, before using a polarizer, it is necessary to be able to verify it. In this way, we propose here a method to evaluate the degree of polarization of a real polarizer, in the circumstances when we haven't any perfect polarizer.

## 2. The degree of polarization of a real polarizer

The experimental arrangement used to this aim is composed of an optical bench where are disposed, in order, a source of natural light (an electric bulb), two identical real polarizing filters (conventionally called, the first – polarizer, and the second – analizer) and a photocell to measure the light intensity.

We make the assumption that the polar diagram of the intensity of the transmitted light by the real polarizer is an ellipse, whom great axis, i.e. the transmission axis of the polarizer, is along Ox axis (fig.2).

Each vibration in the incident natural light having the intensity $I_0$, can be resolved into two components on the axes Ox and Oy of the polarizer. The resultants on each axis have the same intensity, $I_x=I_y=I_0/2$. By passing through the real polarizer, which has the degree of polarization $P$, we admit, for the sake of simplicity, that the component along Ox axis passes not being attenuated, $I_{xp} = I_x$, but the component along the Oy axis is attenuated. So, from the definition (1), one obtains:

$$P = \frac{I_{xp} - I_{yp}}{I_{xp} + I_{yp}} \quad , \tag{2}$$

where the subscripts *xp* and *yp* refer to the components along the respective axes transmitted by the polarizer. From (2) one can obtains the attenuation ratio for Oy direction:

$$\eta = \frac{I_{yp}}{I_y} = \frac{I_{yp}}{I_x} = \frac{I_{yp}}{I_{xp}} = \frac{1-P}{1+P} \tag{3}$$

If the light that passed through the polarizer meets an analyzer, *identical* with the polarizer and having the axes parallel to those of the polarizer, the component parallel to the Ox axis of the analyzer passes unaffected, $I_{xa} = I_{xp}$, while the component on the Oy axis of the analyzer is attenuated by ratio:

$$\eta = \frac{I_{ya}}{I_{yp}} = \frac{1-P}{1+P} \quad . \tag{4}$$

So, behind the analizer the light intensity is maximum and equal to:

$$I_M = I_{xa} + I_{ya} = I_{xp} + \left(\frac{1-P}{1+P}\right)^2 I_{xp} = 2\frac{P^2}{(1+P)^2} I_{xp} \tag{5}$$

Now, if we rotate the analizer by $90^0$ around the light beam (fig. 3), then behind the analizer the two components will have the intensities:

$$I'_{xa} = I_{yp} \quad , \tag{6}$$

because on his Ox axis the analyzer doesn't attenuate, and

$$I'_{ya} = \frac{1-P}{1+P} I_{xp} \quad , \tag{7}$$

the last relation being obtained taking into account of the attenuation ratio corresponding to Oy axis (see (3)).

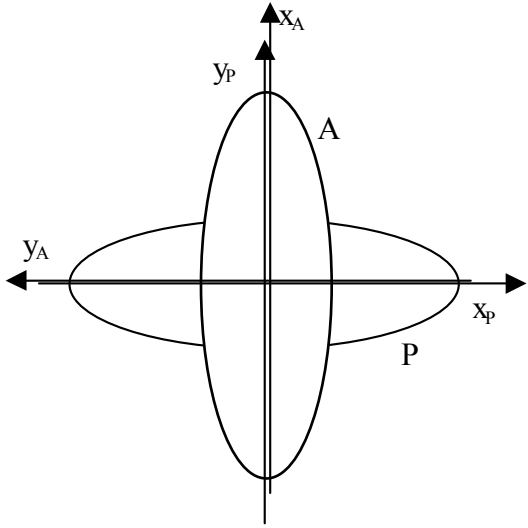 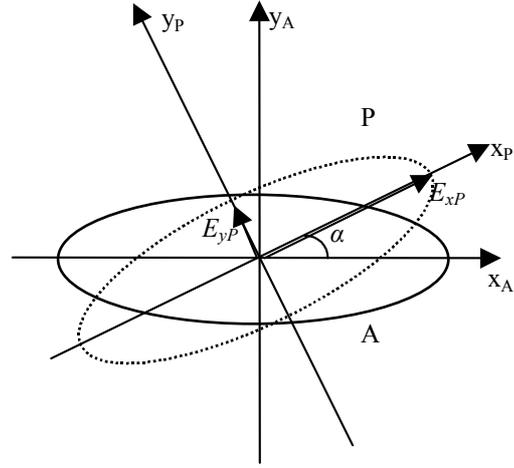

Fig. 3        Fig. 4

So the light intensity behind the analizer will be minimum and equal to:
$$I_m = I'_{xa} + I'_{ya} = I_{yp} + \frac{1-P}{1+P} I_{xp} = 2\frac{1-P}{1+P} I_{xp} \quad (8)$$

Dividing (5) by (8) one obtains:
$$\frac{I_M}{I_m} = \frac{1+P^2}{1-P^2} \quad (9)$$

from which it results the degree of polarization:
$$P = \sqrt{\frac{I_m - I_m}{I_M + I_m}} \quad (10)$$

Thus, measuring the maximum and minimum intensities transmitted by analyzer, from (10) one can obtain the degree of polarization. When the polarizers are perfect, i.e. $P=1$, then from (5) it results $I_M = I_{xp} = I_x$ and from (8) one obtains $I_m = 0$, as it was expecting.

### 3. Malus' law for real polarizer

It is known that Malus' law refers to the beam intensity that passes a *perfect* analyzer, when the incident light is *total* polarized; the law expresses:
$$I = I_0 \cos^2 \alpha \quad (11)$$

where $I_0$ is the linearly (total) polarized incident light intensity, obtained using a perfect polarizer, and $\alpha$ is the angle between the plane of polarization of the light and the transmission axis of the analyzer.

In the case of real polarizers, relation (11) does not hold. In order to establish an adequate relation, let's consider that the $Ox_P$ axis of a real polarizer makes the angle $\alpha$ with $Ox_A$ axis of a real analyzer (fig. 4). The partially polarized light that passes the polarizer is equivalent to two components which oscillate along $Ox_P$ and $Oy_P$ axes, to whom correspond the amplitudes $E_{xP}$ and $E_{yP}$, respectively the intensities $I_{xP}$ and $I_{zP}$. Each of them will have projection onto $Ox_A$ and $Oy_A$ axes, and the corresponding intensities of these projections are:

- for $E_{xP}$:
$$I_{xP}^{(x_A)} = E_{xP}^2 \cos^2 \alpha = I_{xP} \cos^2 \alpha \quad (12)$$
$$I_{xP}^{(y_A)} = E_{xP}^2 \sin^2 \alpha = I_{xP} \sin^2 \alpha \quad (13)$$

- for $E_{yP}$:
$$I_{yP}^{(x_A)} = E_{yP}^2 \sin^2 \alpha = I_{yP} \sin^2 \alpha \quad (14)$$
$$I_{yP}^{(y_A)} = E_{yP}^2 \cos^2 \alpha = I_{yP} \cos^2 \alpha \quad . \quad (15)$$

So, the *incident* light on the analyzer, corresponding to vibrations parallel to his axes, will have the intensities:

$$I_P^{(x_A)} = I_{xP}^{(x_A)} + I_{yP}^{(x_A)} = I_{xP}\cos^2\alpha + I_{yP}\sin^2\alpha \tag{16}$$

$$I_P^{(y_A)} = I_{xP}^{(y_A)} + I_{yP}^{(y_A)} = I_{xP}\sin^2\alpha + I_{yP}\cos^2\alpha \tag{17}$$

Passing through the real analyzer, the light intensity which corresponds to vibrations along $Ox_A$ axis remains unmodified but that which corresponds to vibrations along $Oy_A$ axis is attenuated by a factor $\eta$; thus in view of (16), (17) and (3), the emergent intensity is:

$$I_A = I_P^{(x_A)} + \eta I_P^{(y_A)} = (1+\eta^2)I_{xP}\cos^2\alpha + 2\eta I_{xP}\sin^2\alpha \tag{18}$$

If we take into account that the maximum intensity resulting from polarizer-analyzer assembly is (see (5))

$$I_M = (1+\eta^2)I_{xP} \tag{19}$$

then Eq. (18) reads:

$$I_A = I_M\cos^2\alpha + I_M\frac{2\eta}{1+\eta^2}\sin^2\alpha \tag{20}$$

or, in terms of the degree of polarization, taking into account (3), Eq. (20) becomes:

$$I_A = I_M\cos^2\alpha + I_M\frac{1-P^2}{1+P^2}\sin^2\alpha \tag{21}$$

The relations (20) and (21) represent a generalization of Malus'law; if the polarizers are perfect, $P=1$, then from (21) one obtains the usually form of the Malus' law (11).
The emergent light intensity from analyzer, $I_A$, can be expressed as a function of the incident light intensity, *partially polarized* by polarizer:

$$I_P = I_{xP} + I_{yP} = (1+\eta)I_{xP} = \frac{1+\eta}{1+\eta^2}I_M \tag{22}$$

The relations (20) and (21) become:

$$I_A = \frac{1+\eta^2}{1+\eta}I_P\cos^2\alpha + \frac{2\eta}{1+\eta}I_P\sin^2\alpha \tag{23}$$

and respectively:

$$I_A = \frac{1+P^2}{1+P}I_P\cos^2\alpha + (1-P)I_P\sin^2\alpha \tag{24}$$

which gives the relation between the emergent intensity from analyzer and the incident intensity, similarly to (11).
In Fig. 5 we present, both, the *α*-angle dependence of the experimentally measured emergent intensity from analyzer, using the experimental device described above, and the calculated intensity using (21), in which the polarization degree was determined with (10). The satisfactory concordance of the two curves confirms the correctness of the assumptions of the model used in obtaining of the generalized Malus' law.

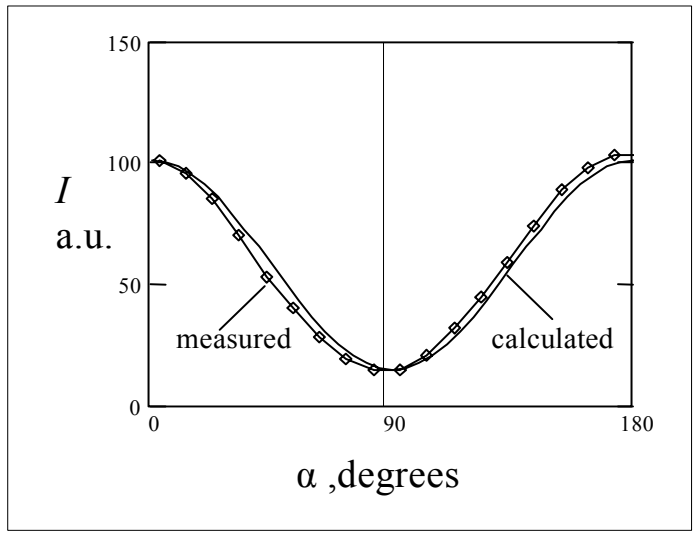

Fig. 5.

## 4. Conclusions

The formula (10), established in the paper, offers the possibility to measure the degree of polarization caused by real polarizers, when we have not any perfect polarizer. The generalized formula of Malus' law supplies the values of the intensity of emerging light from a real polarizer, when the incident light is partially polarized, in good agreement with measured data.

**References**
1. Sears F.W., Zemansky M.W., Young H.D., *Fizica*, Editura Didactica si Pedagogica, Bucuresti, 1983 (tr. from English).